\documentclass[useAMS]{mn2e}

\usepackage{amsmath}
\usepackage{amssymb,graphicx}
\usepackage{verbatim}
\usepackage{journal_names}

\title[A Far-UV Variability Survey of the Globular Cluster M\,80]
{A Far-UV Variability Survey of the Globular Cluster M\,80}
\author[G.~S.~Thomson et al.]
{G.~S.~Thomson,$^1$ A.~Dieball,$^1$ C.~Knigge,$^1$ K.~S.~Long$^2$ and D.~R.~Zurek$^3$ \\
$^1$Department of Physics and Astronomy, University of Southampton, SO17~1BJ, UK\\
$^2$Space Telescope Science Institute, Baltimore, MD 21218\\
$^3$Department of Astrophysics, American Museum of Natural History, New York, NY 10024}

\begin{document}
\pagerange{\pageref{firstpage}--\pageref{lastpage}} \pubyear{2010}
\maketitle
\label{firstpage}
\begin{abstract}
We have searched for variable sources in the core region of M\,80, using far ultra-violet data taken with the Advanced Camera for Surveys on board the Hubble Space Telescope. We found three sources that exhibit strong signs of variability in our data. Among these is source TDK\,1, which we believe to be an RR\,Lyrae star that reached maximum brightness during our observations. The light curve shows a $\gtrsim3$ mag FUV brightening over the course of $\approx5$ hours, with an estimated peak brightness of $\approx16.7$ mag, followed by a decrease to $\approx20$ mag. Archival optical data obtained with WFPC2 confirm that TDK\,1 is variable in all wavebands. TDK\,1's SED is reasonably fit by a star with temperature $T_{eff}\simeq6700$\,K and radius $R\simeq4.2\,R_{\odot}$, consistent with the suggestion that it is an RR\,Lyrae. Based on the photometric and variability characteristics of the other two variables, we suggest that TDK\,2 is likely to be an SX\,Phoenicis star with $\simeq55$ minutes period, and TDK\,3 is likely another RR\,Lyrae. Finally, we briefly discuss the FUV counterparts to two previously known variables in M\,80, the classical nova T\,Sco and the dwarf nova DN\,1.
\end{abstract}


\section{Introduction}

M\,80 is one of the densest Globular Clusters (GCs) in the Milky Way, and is famous for the classical nova T~Scorpii which was discovered in 1860 when it outshone the rest of the cluster. Despite the many observations that resulted from the discovery of the nova, only a few variable sources are known in M80 (Wehlau 1990, Clement, private communication, Kopacki 2009).

Here, we report the results of a variability study based on our far ultra-violet (FUV) survey of M\,80 (Dieball et al. 2010; henceforth Paper~1). We have discovered three stars which exhibit significant variability, the most interesting of which is an RR\,Lyrae star in the core of the cluster. This star, which we call TDK\,1, was observed around the peak of the light curve in our observations, manifesting a high-amplitude ($> 3$ mag), luminous ($L_{\rm{UV}}\simeq6\times10^{34}\ \rm{erg}\ \rm{s}^{-1}$), short-duration ($t\lesssim5\ \rm{hours}$) FUV brightening. Further investigation using archive data show that it is also variable in optical wavebands but the data coverage is not sufficient to determine the period of the variation. The other two new variables discovered in our survey include another possible RR\,Lyrae and an SX\,Phoenicis star. Finally, we discuss the FUV counterparts to two known variables: the famous nova T~Scorpii and the known dwarf nova DN\,1 (Shara \& Drissen 1995).

Our paper is structured as follows: In Section~\ref{obs}, we describe the observations and data analysis, including the FUV survey and analysis of archive HST observations of TDK\,1. The variable sources identified in the cluster are discussed in Section~\ref{variables}. Our results are summarized in Section~\ref{summary}.

\section{Observations and Data Analysis}
\label{obs}
\subsection{UV Data}
\label{FUV_survey_obs}

Our UV survey of the core of M80 consisted of seven orbits of HST observations using the Advanced Camera for Surveys (ACS). Six orbits were spent on FUV imaging, using the F165LP (four consecutive orbits resulting in 32 individual images), the F150LP (one orbit corresponding to eight images) and F140LP (one orbit, eight images) filters in the Solar Blind Channel (SBC). The remaining orbit was used for near ultra-violet (NUV) imaging, using the F220W filter in the High Resolution Channel (HRC) and resulting in eight individual images. All observations were taken at a single pointing position. Table~\ref{data} gives an overview of the UV data.

\begin{table}
\caption{UV and optical data used to study the variable source TDK\,1. The first column gives the camera used, followed by the filter (col.~2), the number of images taken (col.~3), the date of observation (col.~4) and the individual exposure time (col.~5).}
\label{data}
\begin{center}
\begin{tabular}{cccccc}
\hline
\hline
camera & filter & \#  & obs. date  & exp. time\\
       &        &     &            & [sec]\\
\hline
ACS/SBC & F165LP & 8  & 2004-09-11 & 310\\
ACS/SBC & F165LP & 24 & 2004-09-12 & 323\\
ACS/SBC & F150LP & 8  & 2004-09-20 & 310\\
ACS/SBC & F140LP & 8  & 2004-09-26 & 310\\
ACS/HRC & F250W  & 8  & 2004-10-07 & 298\\
WFPC2   & F439W  & 4  & 1994-10-05 & 300\\
WFPC2   & F439W  & 2  & 1996-04-05 & 30\\
WFPC2   & F555W  & 4  & 1996-04-05 & 23\\
WFPC2   & F656N  & 3  & 1997-08-29 & 1300\\
WFPC2   & F675W  & 3  & 1997-08-29 & 260\\
\hline
\end{tabular}
\end{center}
\end{table}

In order to search for variable sources, we used the 32 F165LP images, obtained within a 5.5 hour period. We performed photometry on each individual image, using the FUV source catalogue (see Paper~1) as input to {\tt daophot} (Stetson 1991) running under {\tt IRAF}\footnote{{\tt IRAF} (Image Reduction and Analysis Facility) is distributed by the National Astronomy and Optical Observatories, which are operated by AURA, Inc., under cooperative agreement with the National Science Foundation.}. We used the same parameters as described in Paper~1, but kept the position of the input coordinates fixed (rather than using recentring routines; see Paper 1). For each source we then compared the reduced $\chi^{2}$ values derived using all 32 input images against the mean FUV magnitudes. We found that three sources show a $\chi^{2}$ significantly larger than other sources of comparable brightness, indicating that they are variable. One of these, source no.~2817 in Paper~1, which we henceforth call TDK\,1, exhibited such strong and unusual variability in the FUV that we decided to further investigate this source (see Section \ref{TDK1_other_wavelengths}). The other two sources, source no.~2238 (TDK\,2) and source no.~2324 (TDK\,3), will be discussed in Section~\ref{other}, along with two previously known cataclysmic variables - the classical nova T\,Sco (Pogson 1860) and the dwarf nova DN\,1 (Shara \& Drissen, 1995) - that did not exhibit detectable short term variability in our FUV survey. Table \ref{positions} gives the position and magnitudes of the sources discussed in this paper.

\begin{table*}
\caption{Positions of the variable sources TDK\,1, 2 and 3, as well as DN\,1 and T\,Sco. The first column is the source name, followed by the FUV id number in col.~2. Cols.~3 and 4 give the source position in RA and DEC coordinates. Cols.~5 to 8 give the FUV and NUV magnitudes and corresponding photometric errors as derived from {\tt daophot}. Col.~9 gives the id number of the optical counterpart taken from Piotto et al.~(2002), followed by the optical magnitudes (as reported by Piotto et al.~(2002)) in cols.~10 and 11. Note that the magnitudes given are calculated from the master image in each filter and are thus subject to variability intrinsic to the source.}
\label{positions}
\begin{tabular}{ccccccccccc}
\hline
\hline
1 & 2 & 3 & 4 & 5 & 6 & 7 & 8 & 9 & 10 & 11\\
ID & ID$_{FUV}$ & RA & DEC & FUV & $\Delta$FUV & NUV & $\Delta$NUV & ID$_{piotto}$ & B$_{piotto}$ & V$_{piotto}$\\
 & & [hh:mm:ss] & [deg:mm:ss] & [mag] & [mag] & [mag] & [mag] & & [mag] & [mag]\\
\hline
1 & 2817 & 16:17:02.104 & -22:58:29.23 & 18.209 & 0.014 & 18.045 & 0.005 & 2710 & 16.217 & 16.360\\
2 & 2238 & 16:17:02.164 & -22:58:33.24 & 19.981 & 0.032 & 19.703 & 0.011 & 2172 & 19.015 & 19.028\\
3 & 2324 & 16:17:02.861 & -22:58:32.65 & 18.593 & 0.016 & 17.403 & 0.003 & 1450 & 16.594 & 16.226\\
DN\,1 & 1387 & 16:17:02.176 & -22:58:38.59 & 22.578 & 0.120 & 21.869 & 0.055 & * & * & *\\
T\,Sco & 2129 & 16:17:02.818 & -22:58:33.94 & 15.444 & 0.005 & 19.247 & 0.008 & * & * & *\\
\hline
\end{tabular}
\end{table*}

\begin{figure*}
\includegraphics[height=6cm]{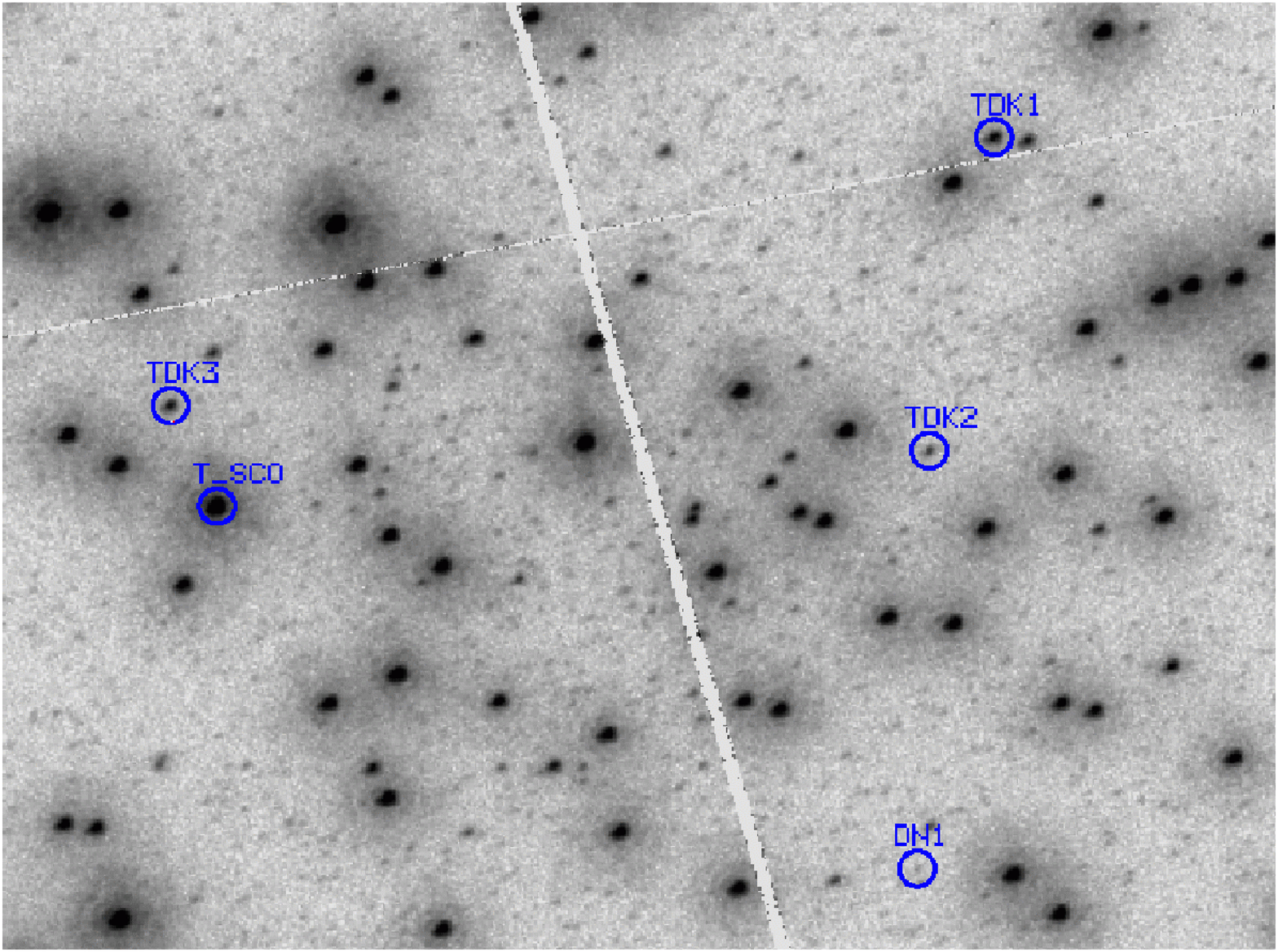}
\includegraphics[height=6cm]{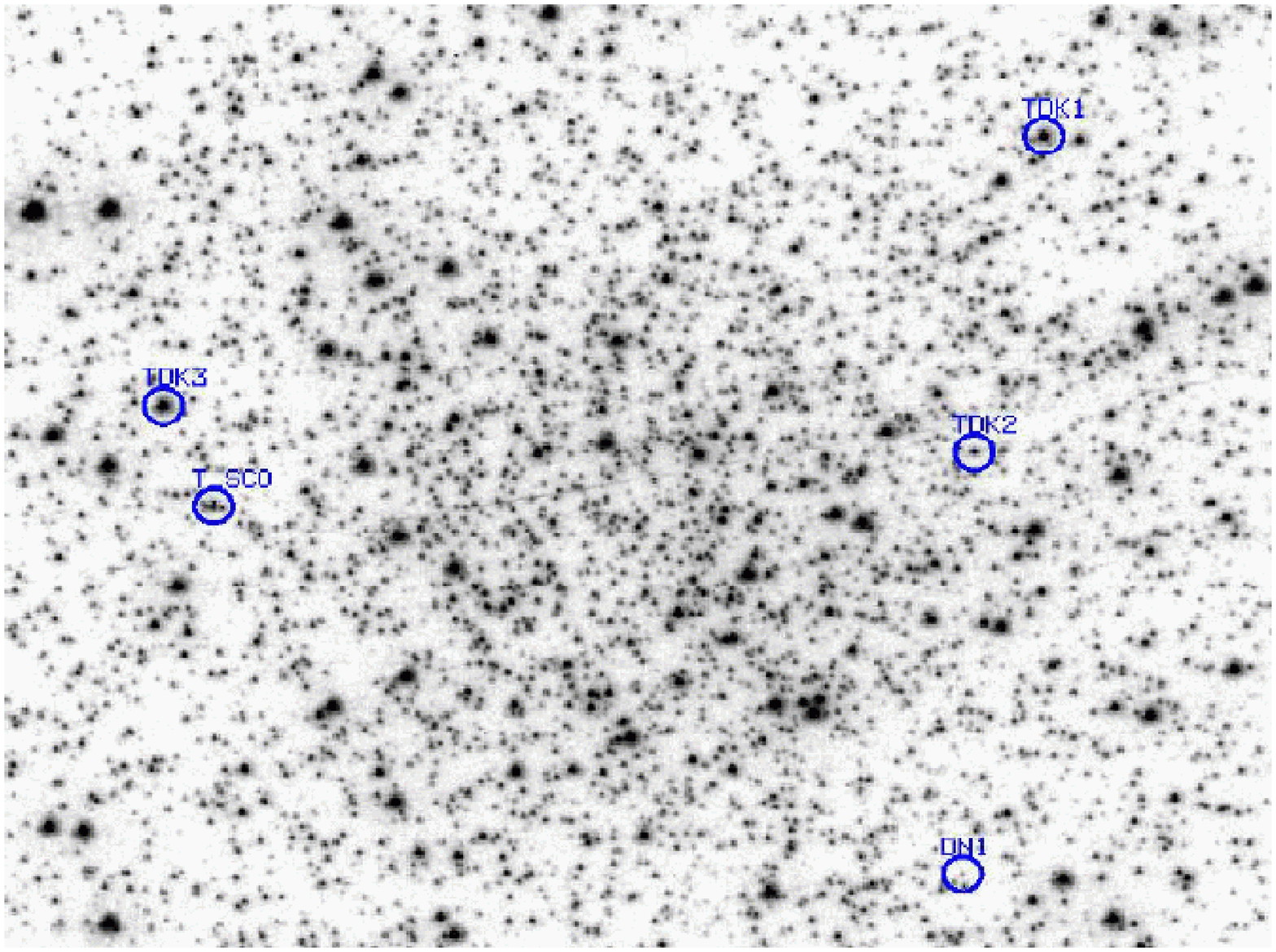}
\caption[finder_all]{Portion of the combined FUV (left) and NUV (right) image showing the five variable sources discussed in this paper. North is up and East is to the left.}
\label{fig_finder_all}
\end{figure*}

In Figure~\ref{fig_finder_all}, the parts of the master FUV and NUV images that include the five sources discussed in this paper are shown.

\subsection{Optical Observations of TDK\,1}
\label{TDK1_other_wavelengths}

\begin{figure*}
\includegraphics[height=4cm]{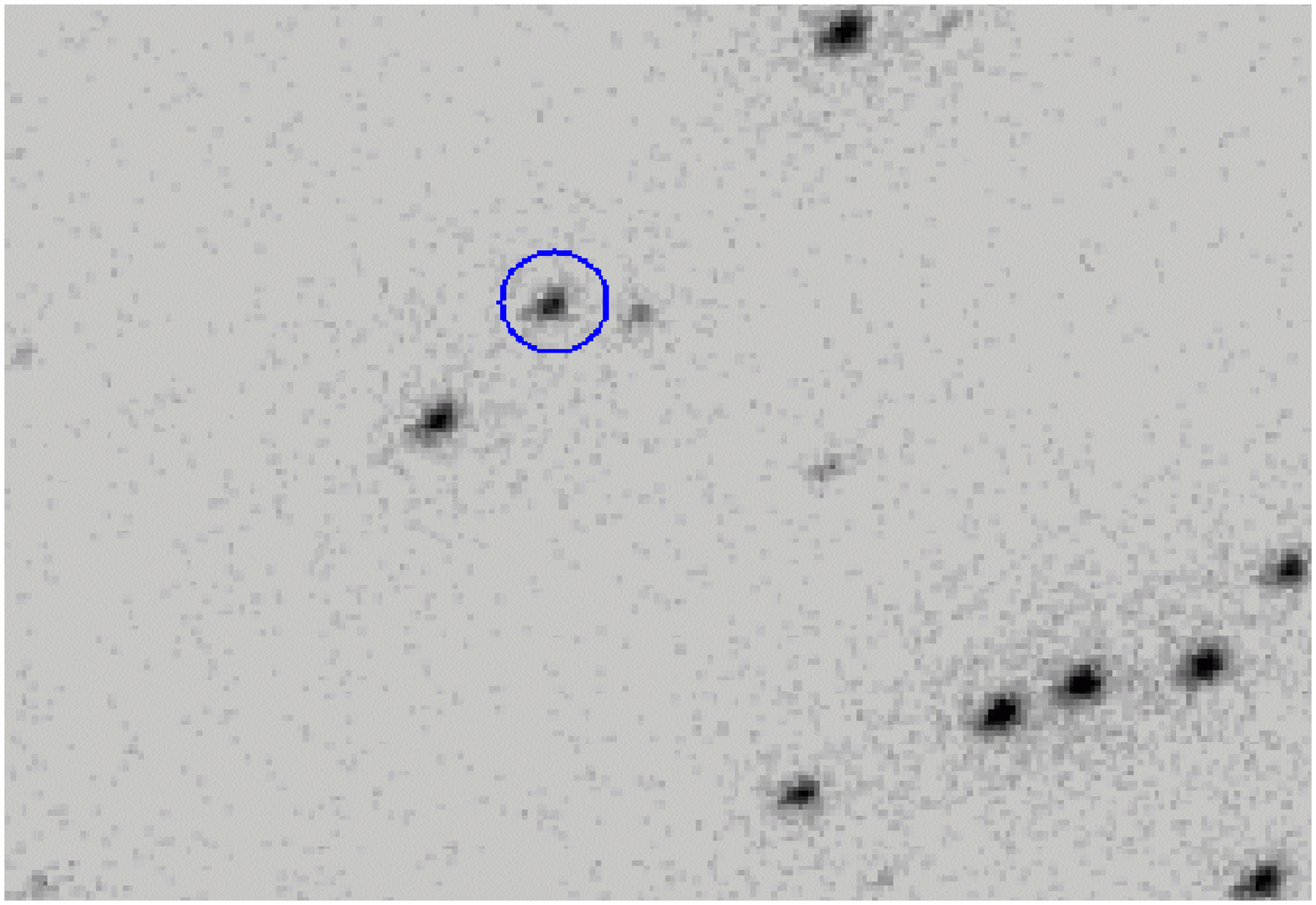}
\includegraphics[height=4cm]{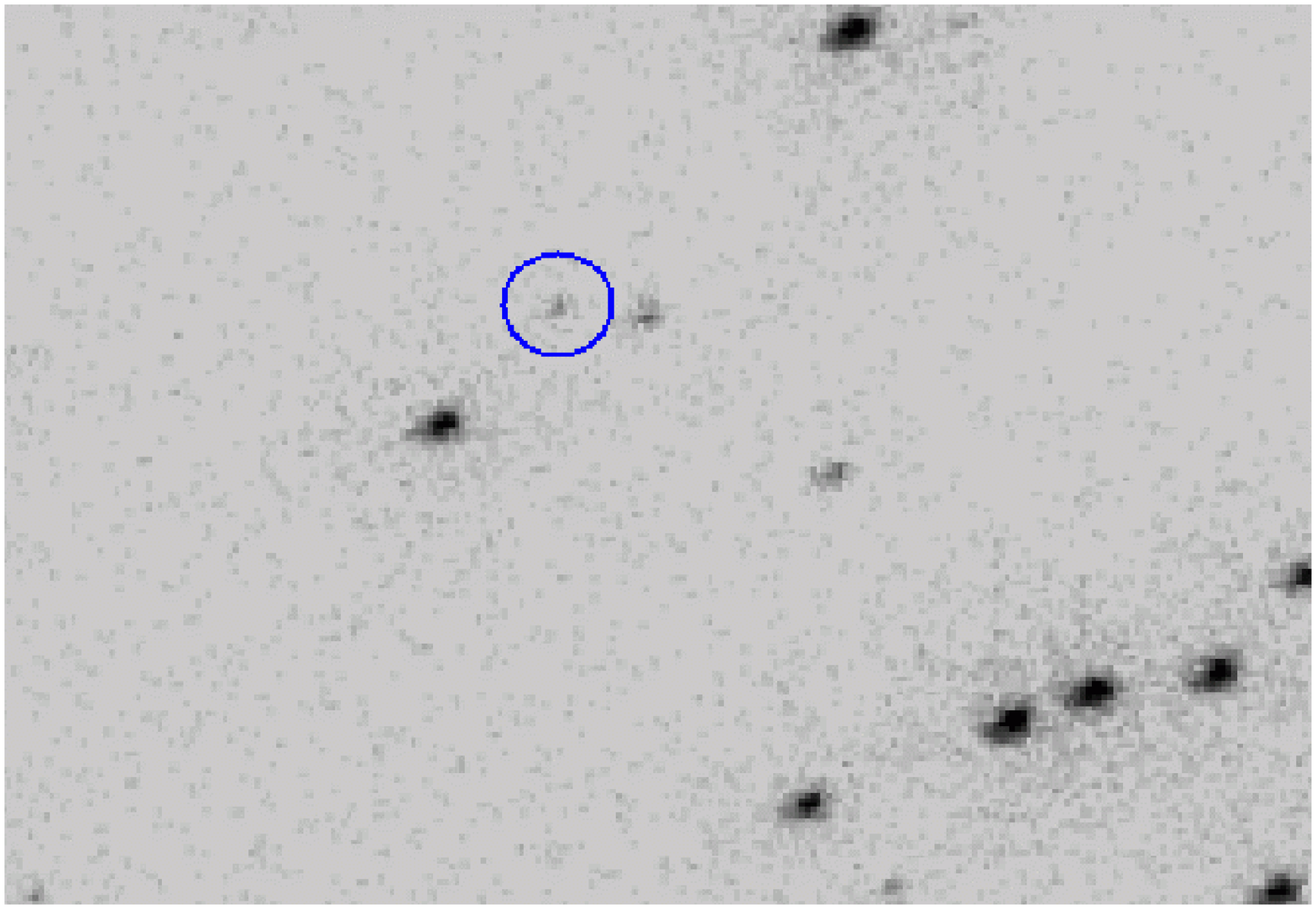}
\caption[tdk_finder_fuv]{The variable source, TDK\,1, at the brightest (left) and the faintest (right) that was seen in our FUV observations. North is up and East is to the left. Note that the difference in brightness shown between the left and right images is an underestimate of the actual variation: The high state magnitude is an underestimate of the peak brightness, due to a gap in the data at this point; the low state magnitude is not the true low state, as the source was still growing fainter when this observation occurred. See text for details.}
\label{outburst}
\end{figure*}

\begin{figure*}
\includegraphics[height=4cm]{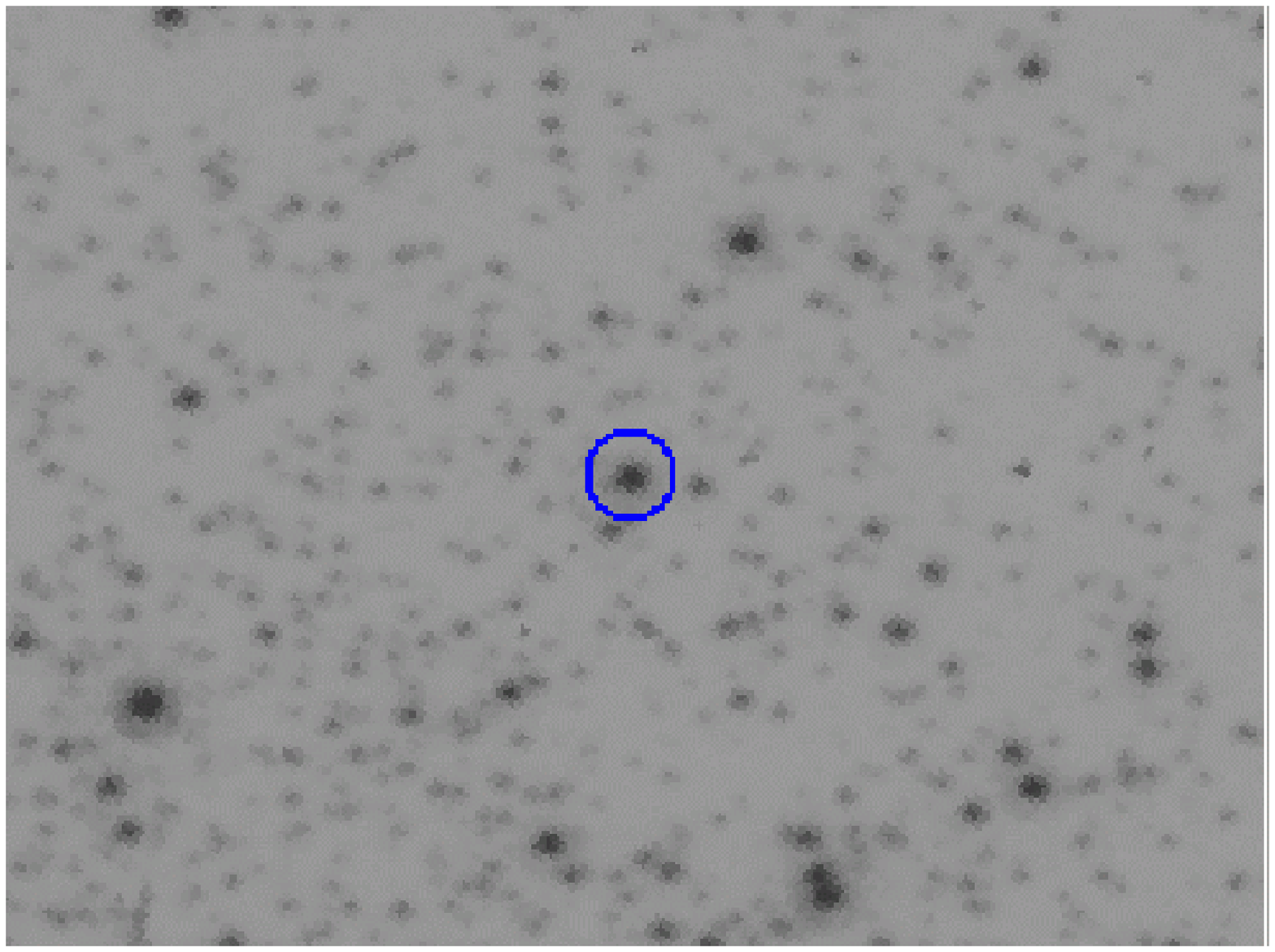}
\includegraphics[height=4cm]{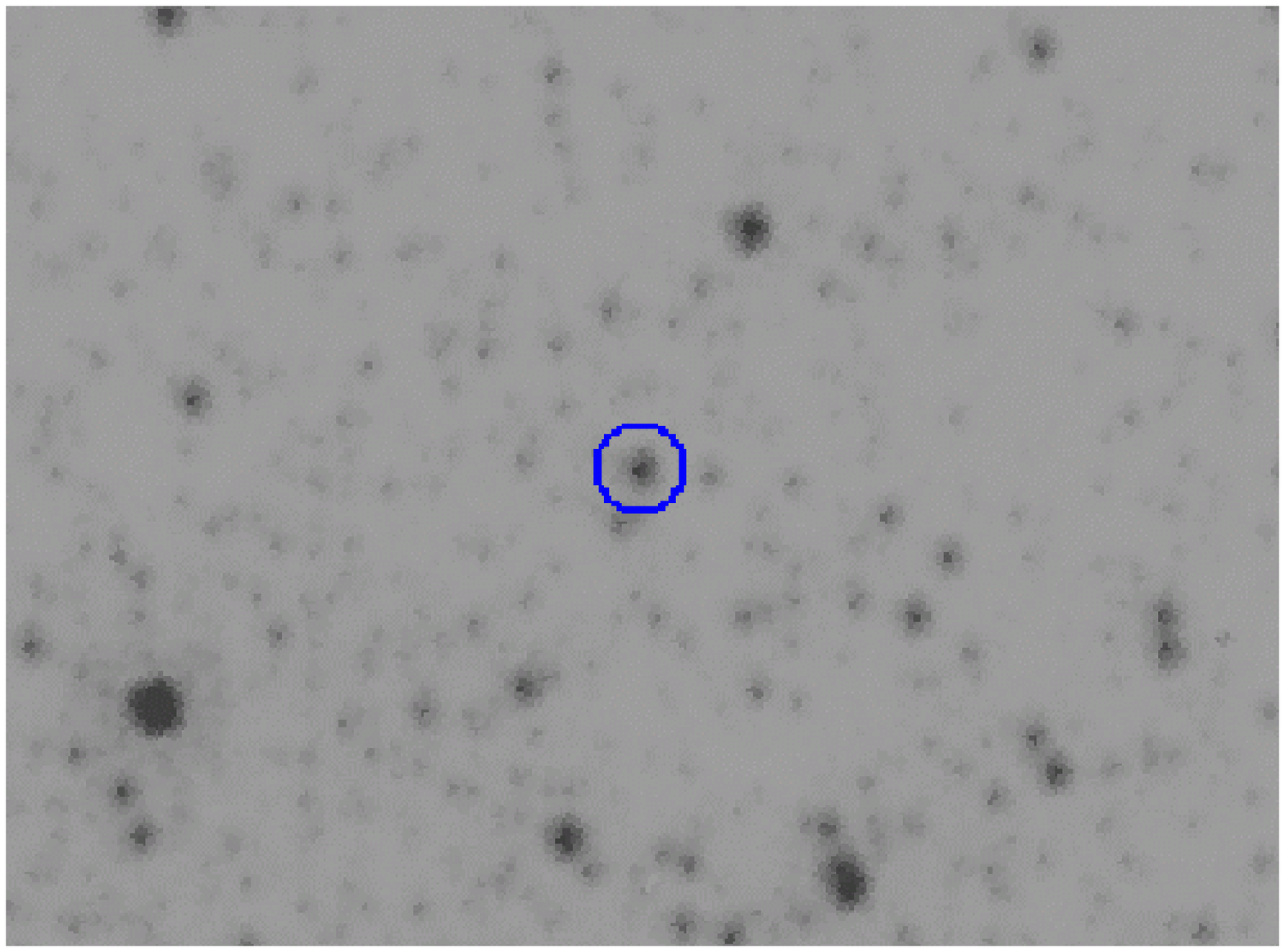}\\
\includegraphics[height=4cm]{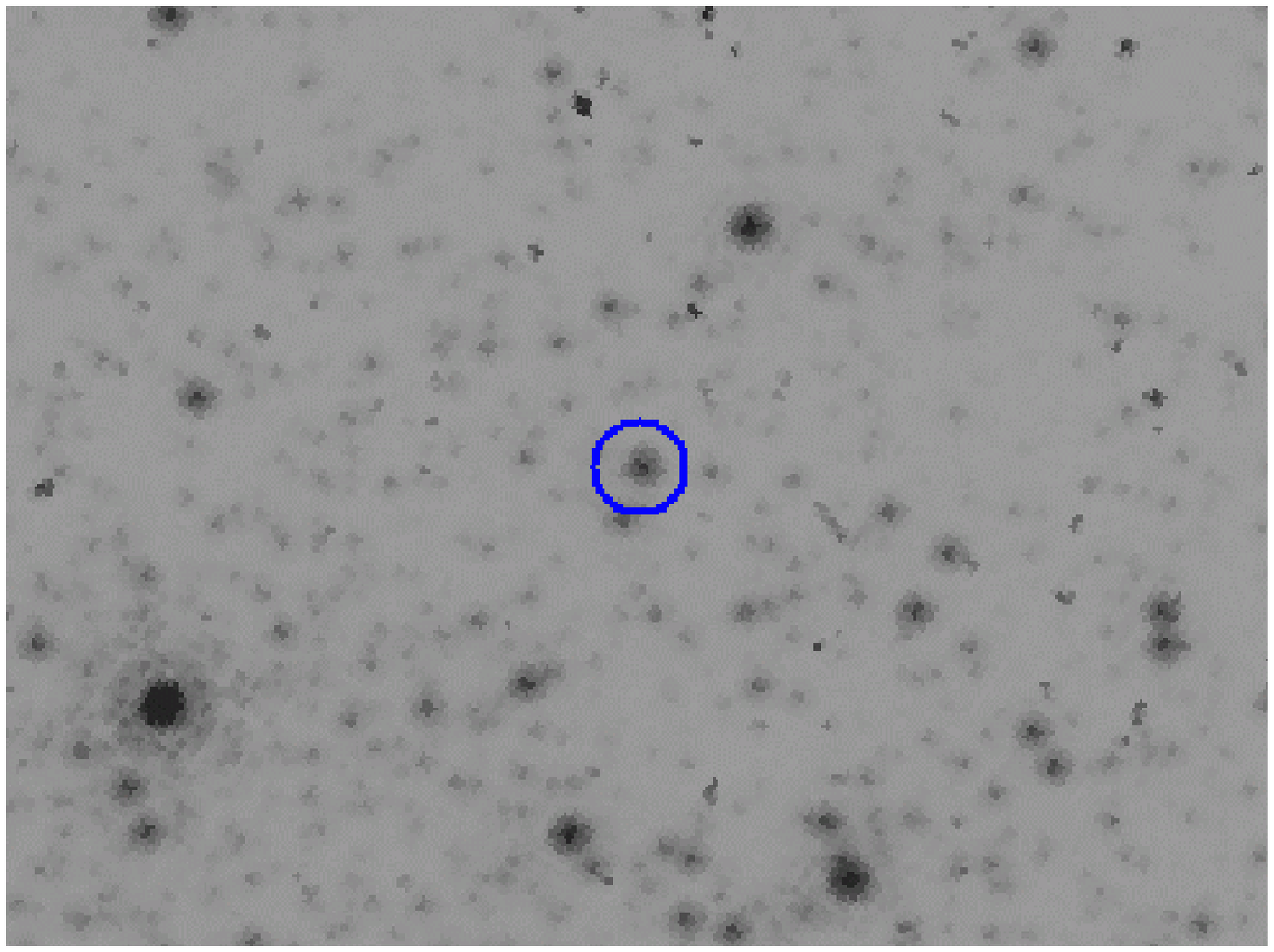}
\includegraphics[height=4cm]{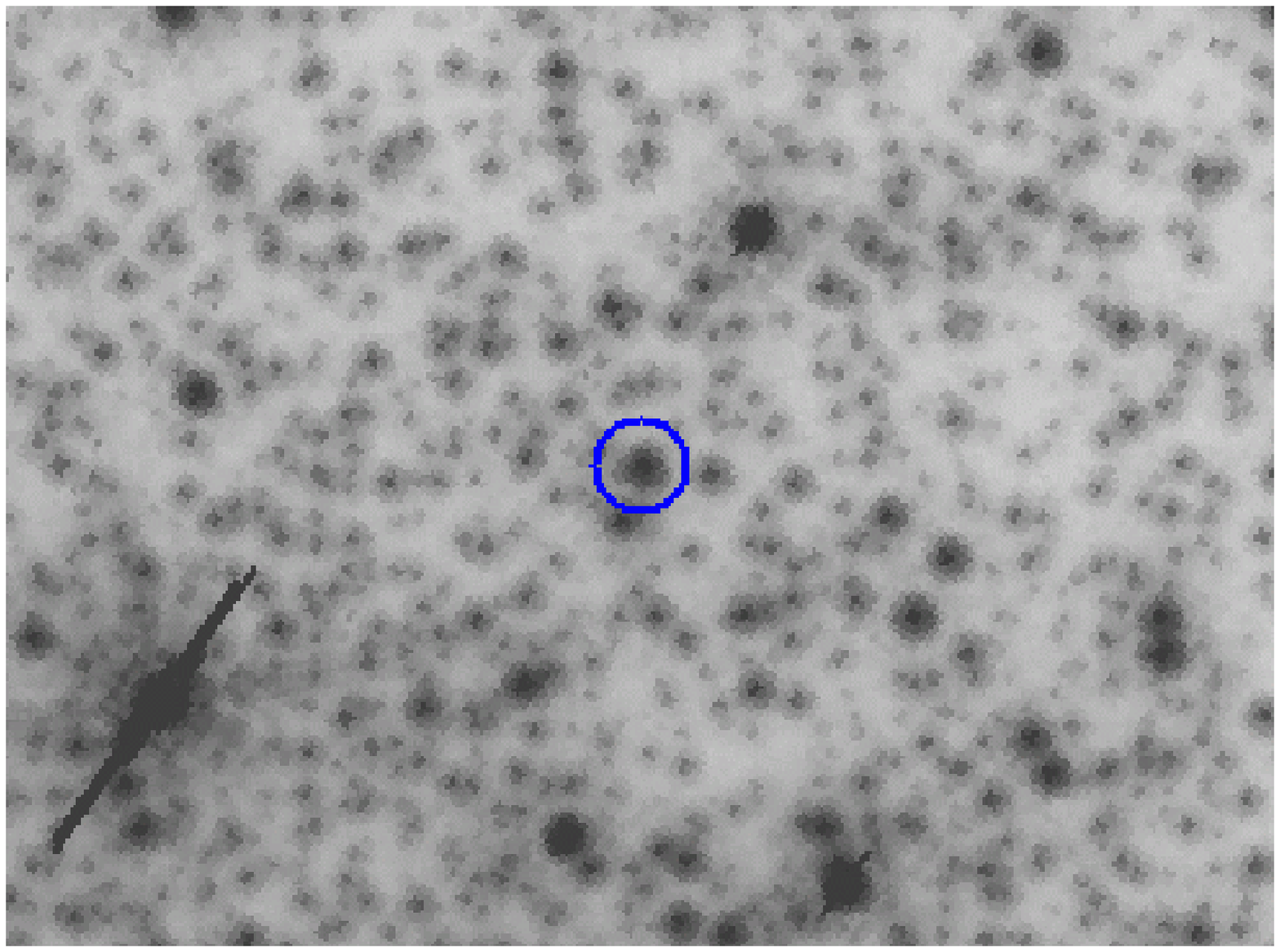}
\caption[tdk_finder_all]{The variable source, TDK\,1, as seen in the F439W (top left), F555W (top right), F656N (bottom left) and F675W (bottom right) filters. North is up and East is to the left.}
\label{fig_finder_filters}
\end{figure*}

As mentioned above, TDK1 is highly variable, brightening by more than 3 magnitudes during our observations. Figure~\ref{outburst} shows the FUV image of TDK\,1 when it is brightest (left) and faintest (right). In order to gather more information about this source, we analysed additional data of the core region of M\,80, taken with the Wide-Field Planetary Camera 2 (WFPC2), available from the HST archive. In total, we used six images taken with the F439W filter, four images taken in F555W, three images in the narrow-band $H\alpha$ filter F656N, and three images in the broad-band $H\alpha$ filter F675W. Table~\ref{data} gives an overview of all the data used to study TDK\,1, including the filters used, and the observing and exposure times. Figure~\ref{fig_finder_filters} shows TDK\,1 as it appears in individual images taken with each of these other filters.

\begin{figure*}
\includegraphics[width=1\textwidth]{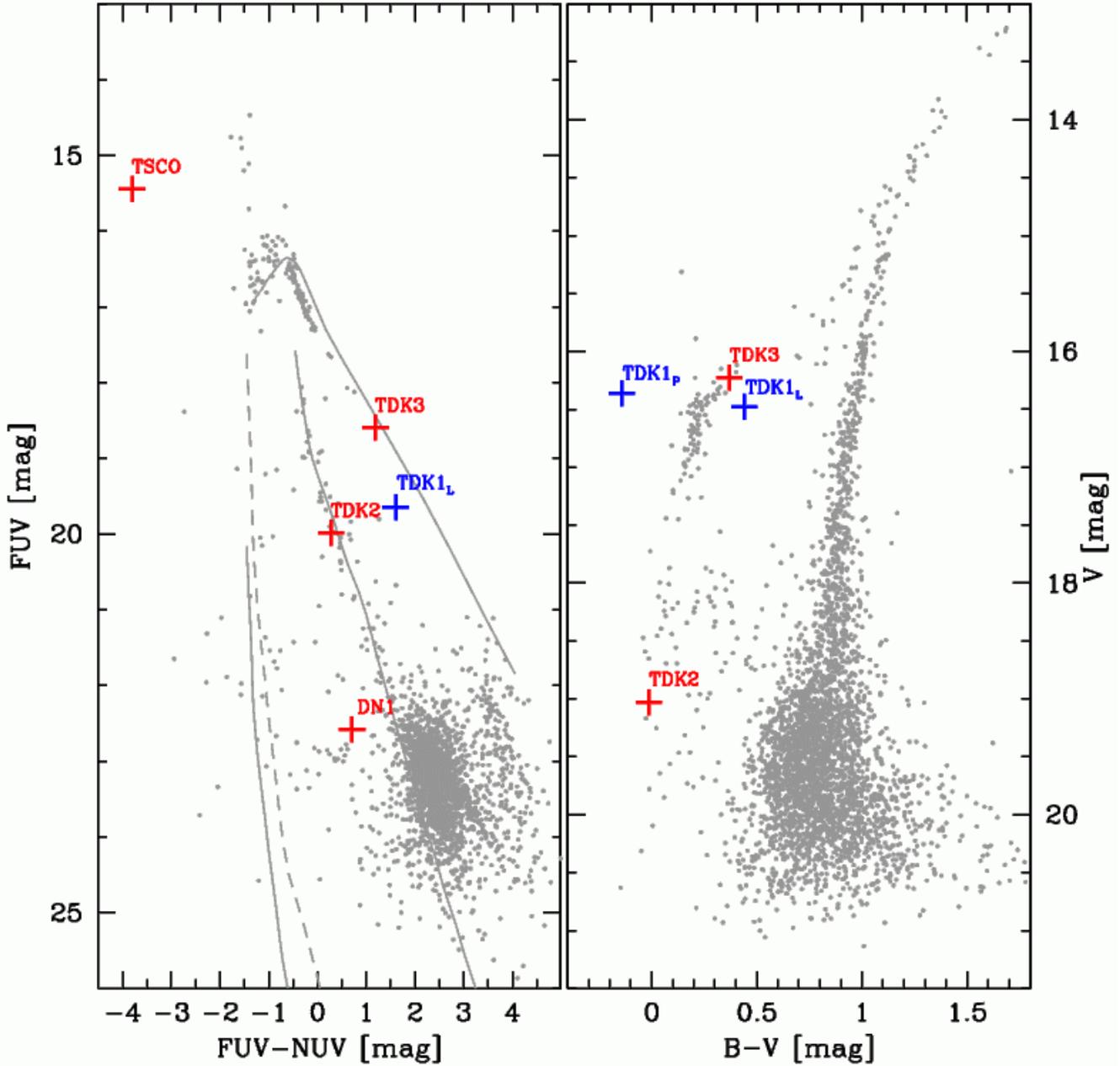}
\caption[CMD]{Left: FUV-NUV CMD of the core of M\,80, with the variable sources marked in red, except for TDK\,1 which exhibited large variation in FUV magnitude and is highlighted in blue. Note that the FUV magnitude for TDK\,1 is the average value from the last eight data points of the F165LP data only, to indicate the minimum brightness location. We caution that this is still an overestimate of the FUV brightness since the system was still declining from its outburst during the last F165LP orbit. All other magnitudes are derived from the master image of the given filter (F165LP and F250W). For reference, we also include a theoretical WD and He WD cooling sequence (dashed and solid line towards the blue), a zero-age main sequence (middle), and a zero-age HB track (reddest/brightest). See Paper~1 for details. Right: Optical CMD of M\,80 with the optical counterparts of the variable sources marked, as in the Piotto et al. (2002) catalogue. Piotto et al. used master images created from two F439W and four F555W observations. For TDK\,1 we also include the position obtained using only the faintest data point in each band, which gives, as close as possible with our data, the `low-state' position (see Section \ref{sed} for details). This point is marked TDK\,1$_{L}$, whereas Piotto et al.'s position is TDK\,1$_{P}$.}
\label{fig_cmd}
\end{figure*}

In order to determine whether TDK\,1 is variable in other bands, we performed relative photometry using nine non-varying stars of similar magnitude for comparison. In each data set, the sources were identified and photometry was performed on the individual images. In each case, a 3 pixel aperture and a small sky annulus of $5-7$ pixels was used. Although the core of M80 is very crowded (particularly in the optical images), TDK\,1 and the nine stars chosen for comparison are bright and relatively isolated, so an aperture of 3 pixels was a good compromise between trying to include most of the light from the source and avoiding flux from neighbouring stars. The same photometry procedure was followed as for the FUV images, but published aperture corrections from Holtzman et al. (1995) were used. For the narrow-band $H\alpha$ data, the aperture correction is not included in the list published by Holtzman et al. (1995), so we estimated the correction using the values given for the bracketing filters (in wavelength space). The STMAG system is used throughout, except for the optical magnitudes in Table \ref{positions} and Figure \ref{fig_cmd} which are taken from Piotto et al. (2002) and are Johnson magnitudes.

\section{FUV Variable Sources in the Core of M\,80}
\label{variables}

We identified 3 objects in our FUV survey of M80 that exhibit significant, short-term variability. One of these sources, TDK\,1, showed exceptional variability and warranted further investigation. In this section we first discuss the light curves and spectral energy distribution of this source and classify it as an RR\,Lyrae star. We also discuss the other two variable sources found in our FUV variability survey, and two previously known variables, T\,Sco and DN\,1.

\subsection{TDK1}
\label{tdk1}

In this section we will discuss the results of our investigation into TDK\,1 and present evidence that this source is an RR\,Lyrae star.

RR\,Lyrae stars are HB stars with $T_{eff}$ about $6000-7600\,K$ and radii between 4 and 7\,$R_{\odot}$ (e.g., Smith 1995, L\'azaro et al. 2006, Pe\~na et al. 2008, S\'odor et al 2009), which exhibit periodic variability with amplitude $0.2-2$ mag at optical wavelengths and up to 8 mag at UV wavelengths (Wheatley et al. 2005). The variability is well understood to be due to radial pulsations that result in radius and effective temperature changes. Although many of these stars have been observed in near-infrared and optical wavelengths and the form of the light curve in these ranges is well established, very few have been well observed at wavelengths shorter than $\sim 3000$\,\AA. Among the examples of short wavelength observations of RR\,Lyraes, Bonnell \& Bell (1985) and Fernley et al. (1990) observed a selection of RR\,Lyrae stars using the International Ultra-Violet Explorer in the range $2000-3000$\,\AA, and Downes et al. (2004) discovered 11 RR\,Lyrae stars in the core of the GC NGC\,1851 using HST Space Telescope Imaging Spectrograph FUV observations. Wheatley et al. (2005) presented the first observation of an RR\,Lyrae star in which an entire cycle was seen using simultaneous FUV, NUV and optical instruments. The variation in apparent magnitude is much more extreme at shorter wavelengths (Wheatley et al. (2005) predict FUV amplitudes of up to $\sim8$\,magnitudes). This makes FUV observations a potentially useful tool for identifying new RR\,Lyrae (or similar) stars, so the lack of UV observations of RR\,Lyrae stars is surprising. One should note, however, that the FUV magnitudes of RR\,Lyrae stars away from maximum brightness are very faint compared with optical and near-infrared values, so observations long enough to include an entire period or fortuitous observations of maxima would be required to allow such variables to be recognised.

There are two main, distinct groups of RR\,Lyrae: RR\,ab, which are fundamental mode pulsators, usually have periods of $\gtrsim0.4$\,days and are characterised by their asymmetric light curves, and RR\,c, which have first overtone pulsations, and have shorter periods and smaller amplitudes than RR\,ab, with more sinusoidal light curves. The ratio of RR\,ab$:$RR\,c is thought to be related to the metallicity of the cluster. Oosterhoff I (`metal-rich') clusters have $N_{c}/(N_{ab}+N_{c})\sim0.2$ while Oosterhoff II (`metal-poor') clusters have $N_{c}/(N_{ab}+N_{c})\sim0.5$ (Oosterhoff 1939, Bono et al. 1994). Previous studies of variable sources in M\,80 have attempted to classify the cluster according to Oosterhoff's criteria, but small number statistics have meant that it has never been convincingly determined. Until recently, only six RR\,Lyrae were known in M\,80, of which 4 were RR\,ab, making M\,80 a borderline Oosterhoff~II cluster. Kopacki (2009) raised the totals to 7 RR\,ab and 8 RR\,c; with an RR\,c fraction of 53\%, confirming the Oosterhoff~II classification. This is consistent with other classification methods, for example Alcaino et al. (1998) and Cavallo et al. (2004) found an iron abundance of [Fe/H]$\simeq-1.7$.

\subsubsection{Variability}

\begin{figure*}
\includegraphics[angle=270,width=0.95\textwidth]{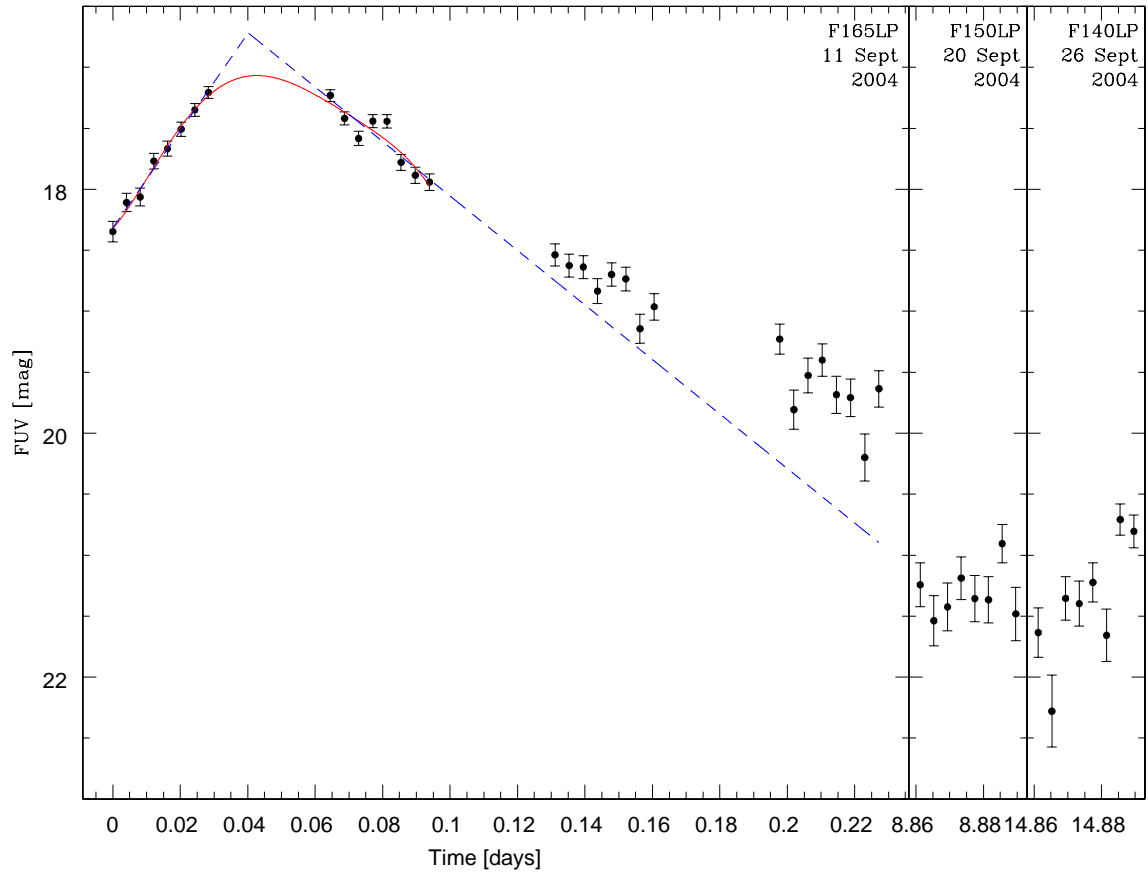}
\caption[Light curve of 2817 in FUV]{Light curve for the variable source TDK\,1, including all FUV data. The lines show simple polynomial fits to the first two orbits of data, illustrating the range of possible peak magnitudes of the outburst. The blue dashed line shows straight line fits to each of the first two orbits' points, while the solid red line shows a 5th order polynomial fit to the data. Left panel: A set of 32 images taken using the F165LP filter. Middle panel: Eight images taken approximately 9 days later using the F150LP filter. Right panel: A set of eight F140LP images taken around 15 days after the first image.}
\label{fig_lightcurve_f165_150_140}
\end{figure*}

The FUV light curve for TDK\,1 is displayed in Figure~\ref{fig_lightcurve_f165_150_140}. The left, middle and right panels show the data obtained with the F165LP, F150LP and F140LP filters, respectively. The star increases in brightness from 18.3 mag to 17.2 mag within the first orbit ($\approx 40$ min) and then fades over the next three orbits ($\approx 4.5$~hours) to $\approx 20$ mag. Due to the gap in the data between the consecutive orbits, the peak in the light curve is not observed. Simple linear fits to the light curve on either side of the peak (shown in Figure \ref{fig_lightcurve_f165_150_140}) suggest that TDK\,1 might have reached FUV $\simeq 16.7$ at peak. The F150LP and F140LP data sets suggest that the source is in a low brightness state at around FUV $\approx 21$ mag, giving a difference between the minimum (F150LP data) and maximum brightness (F165LP data) of around 4 magnitudes. This is consistent with the FUV amplitudes of RR\,Lyraes, as demonstrated by Downes et al. (2004) and Wheatley et al. (2005). Furthermore, the shape of the F165LP light curve is asymmetric - the rise to maximum brightness is considerably steeper than the decline that follows - indicating that TDK\,1 is an RR\,Lyrae star of type~ab. Comparing TDK\,1's light curve with the data given by Wheatley et al. (2005), we estimate that TDK\,1's period is $\approx0.84$~days. This is consistent with the conclusion that TDK\,1 is a type ab RR\,Lyrae.

\begin{table}
\caption[TDK1 in other wavebands]{Magnitude variations in TDK1 compared to the average values for a collection of nine comparison stars.}
\label{TDK1_table}
\begin{tabular}{@{}lcccc}
\hline
\hline
Filter & \multicolumn{2}{c}{RMS Variation [mag]} &
\multicolumn{2}{c}{Peak to Peak Variation [mag]} \\
 & TDK1 & Comparison & TDK1 & Comparison\\
\hline
F439W & 0.24 & 0.02 & 0.62 & 0.07\\
F555W$^a$ & 0.07 & 0.03 & 0.18 & 0.07\\
F656N & 0.11 & 0.01 & 0.26 & 0.02\\
F675W & 0.04 & $<0.01$ & 0.10 & 0.01\\
\hline
\end{tabular}
$^a$ In the F555W filter image, only eight comparison stars were used as the ninth was outside the field of view.
\end{table}

Photometric measurements based on the available HST optical imagery (see Section~\ref{TDK1_other_wavelengths}) show, as might be expected, that TDK\,1 is variable in all wavebands. Table~\ref{TDK1_table} shows the RMS and peak-to-peak magnitude variation for TDK\,1, and the average values for the nine comparison stars. TDK1 exhibits more variability than the comparison stars in all bands.

The variability in all wavebands makes placing TDK\,1 on a colour-magnitude diagram (CMD) something of a challenge. Figure \ref{fig_cmd} shows the CMD of M\,80, with the variable sources highlighted. For TDK\,1, the UV magnitudes are the averages from the orbits in which it was faintest in order to minimise the effect of variability on the CMD position. In the UV CMD, TDK\,1 is located close to the blue horizontal branch (BHB)/blue straggler (BS) region, as might be expected for an RR\,Lyrae, although its position is still subject to error due to variability even excluding the data taken during the peak of the light curve. The optical CMD was created using data from Piotto et al. (2002), in which the average magnitudes from two F439W and four F555W observations were used (observations taken in 1996; see Table \ref{data}); Piotto's location for TDK\,1 is indicated with subscript P. To minimise the effect of variability in these two wavebands, we also plot the `low-state' position of TDK\,1, marked TDK\,1$_{L}$. To do this we used the faintest data point in each band and converted them to Johnson magnitudes using the {\tt IRAF/STSDAS} package {\tt Synphot}. This brings TDK\,1 much closer to the BHB. As explained in more detail in Section \ref{sed}, TDK\,1 was not exactly at minimum brightness in these observations, so this is still not the true low state; furthermore, the F555W data was taken at a brighter point in the cycle than the F439W data, making the source appear redder than if they were taken simultaneously. We conclude that, subject to errors due to variability, TDK\,1's position on both CMDs is consistent with it being an RR\,Lyrae.

\subsubsection{Spectral Energy Distribution}
\label{sed}

In order to confirm that TDK\,1 is an RR\,Lyrae star, as suggested by the variability in the FUV and optical wavebands, we constructed the spectral energy distribution (SED) of TDK\,1. In doing so we tried to use observations taken when the star was close to minimum brightness (`low state'), but we caution that lack of truly low-state observations mean the SED produced is still not actually that of minimum brightness. For the FUV and NUV we used the average magnitude from the faintest orbits; for F165LP we include only the average magnitude from the last eight data points to minimise the effect of the large magnitude variation exhibited in the light curve. We also tried excluding the F165LP data entirely, and found that this did not significantly change the results. For the optical bands we know, as shown in Table \ref{TDK1_table}, that there was some variability present in each dataset. In fact, TDK\,1 is undergoing brightening in the 1996 observations (the last two F439W images, followed by the four F555W images; see Table \ref{data}), and again in the 1997 observations (three F675W observations followed by three F656N observations). To minimise the effect of this variability we used the faintest magnitude measurement for each filter, to give the best possible approximation of a low-state SED. While we expect that this issue will produce errors of a few tenths of a mag in each waveband, it is sufficiently accurate to get an idea of the shape of the SED and to estimate parameters such as temperature and radius.

\begin{figure*}
\includegraphics[angle=270,width=0.95\textwidth]{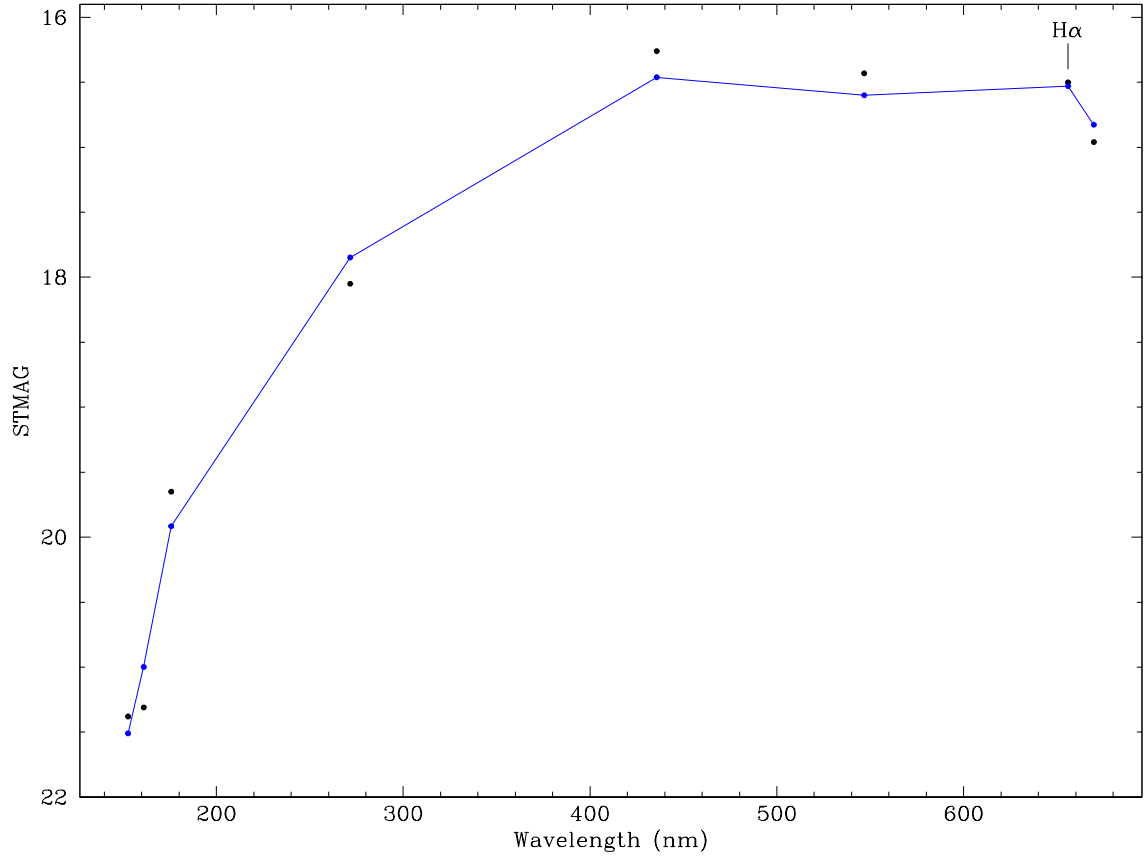}
\caption[SED]{Spectral Energy Distribution (SED) for TDK\,1 (black points), along with the best model fit to the data (blue line), which has $T_{eff}\simeq6700$\,K, $R\simeq4.2\,R_{\odot}$, and log $g\simeq3.0$, leading to a mass estimate of $M\simeq0.6\,M_{\odot}$.}
\label{fig_sed_synphot}
\end{figure*}

TDK\,1's SED is shown in Figure~\ref{fig_sed_synphot} (black data points). Each point is plotted at the pivot wavelength for the corresponding filter. The SED is consistent with a single star. Figure~\ref{fig_sed_synphot} also shows that the source appears brighter in the narrow-band F656N filter than in the bracketing optical broadband data points in the SED, suggesting an $H\alpha$ excess.

Synthetic photometry was carried out for models in the Kurucz (1993) grid using the {\tt IRAF/STSDAS} package {\tt Synphot} and assuming a distance of 10 kpc, a reddening of $E_{B-V}=0.18$ mag and a metallicity of $\rm{[Fe/H]}=-1.75$ dex. The synthetic SEDs were then fit to the data using a least squares fitting method, in which each point was given equal weight. We found that TDK\,1's SED can be reasonably described by a star with $T_{eff}\simeq6700$\,K, $R\simeq4.2\,R_{\odot}$, and log $g\simeq3.0$, giving a mass of $M\simeq0.6\,M_{\odot}$. These values are all within the acceptable range for RR\,Lyrae stars.

In order to estimate the contribution of the $H\alpha$ line to the underlying stellar spectrum, we repeated the synthetic photometry including an $H\alpha$ line of varying equivalent width (EW). The best fit model suggests a marginally significant ($\approx 2 \sigma$) $H\alpha$ line with EW$\simeq20$\,\AA, without significantly affecting the other fit parameters. The best fit SED, including an $H\alpha$ line with EW$\simeq20$\,\AA, is plotted as a blue line in Figure~\ref{fig_sed_synphot}. The presence of an $H\alpha$ line can be explained by the fact that TDK\,1 is undergoing a brightening phase during the F656N and F675W observations. This adds further evidence that TDK\,1 is a type~ab RR\,Lyrae, as these are known to exhibit $H\alpha$ emission during rising light (Smith 1995).

As shown in Figure~\ref{fig_sed_synphot}, the models do not fit well to the observed slope in the optical data; the data in this region shows a much steeper slope (i.e. bluer colour) than the models. As the pivot wavelengths for the optical filters are similar, the colours indicated by comparisons between these filters are sensitive to, and can be easily distorted by, variability. The variability that we observe in the optical filters (a few tenths of a mag) is, therefore, sufficient to account for the very blue optical colour of TDK\,1. Fitting the slope of the optical data alone gives a far higher $T_{eff}$ than our best-fit model. In fact, such a model over predicts the brightness in the UV by such a large amount that it becomes impossible to account for the observed UV magnitudes. Furthermore, it is the data taken using the F140LP and F150LP filters that we trust most to have been taken in the low-state. Thus, $T_{eff}$ can be well determined by ensuring that the model is a good match to the magnitudes obtained using the F140LP and F150LP filters.

It is worth noting that while $T_{eff}$ is well constrained by the turnover in the SED and the radius by the flux normalization to M\,80's distance, the global shape of the SED can be reasonably well described using a range of log $g$ values. Thus, a slight change in input $T_{eff}$ or $R$ has a huge impact on the model SED, whereas large changes in log $g$ only have relatively small effects. RR\,Lyrae stars have been noted to have a wide range of log $g$ values (see Pe\~na et al. 2009 for examples of RR\,Lyraes with log $g$ values of 1.3 to 2.2 at minimum brightness, and 2.5 to 3.5 at maximum), so while our best-fit value of log $g\simeq3$ is reasonable, we caution that variability in the data limits our ability to make a reliable determination of log $g$ and, consequently, the mass. By contrast, the temperature and radius estimates should be fairly reliable.

We conclude, based on the radius and temperature obtained from the SED and the suggestion of H$\alpha$ excess during the brightening phase, that TDK\,1 is indeed an RR\,Lyrae star. The log $g$ and mass values obtained are also consistent with this explanation, but are less reliable as they are affected more significantly by slight variations in magnitude.

\subsection{Other FUV Variable Sources: TDK\,2 and 3}
\label{other}

In this section, we briefly discuss the nature of two other sources in our catalogue that showed strong signs of FUV variability.

\subsubsection{TDK\,2}

\begin{figure*}
\includegraphics[angle=270,width=1\textwidth]{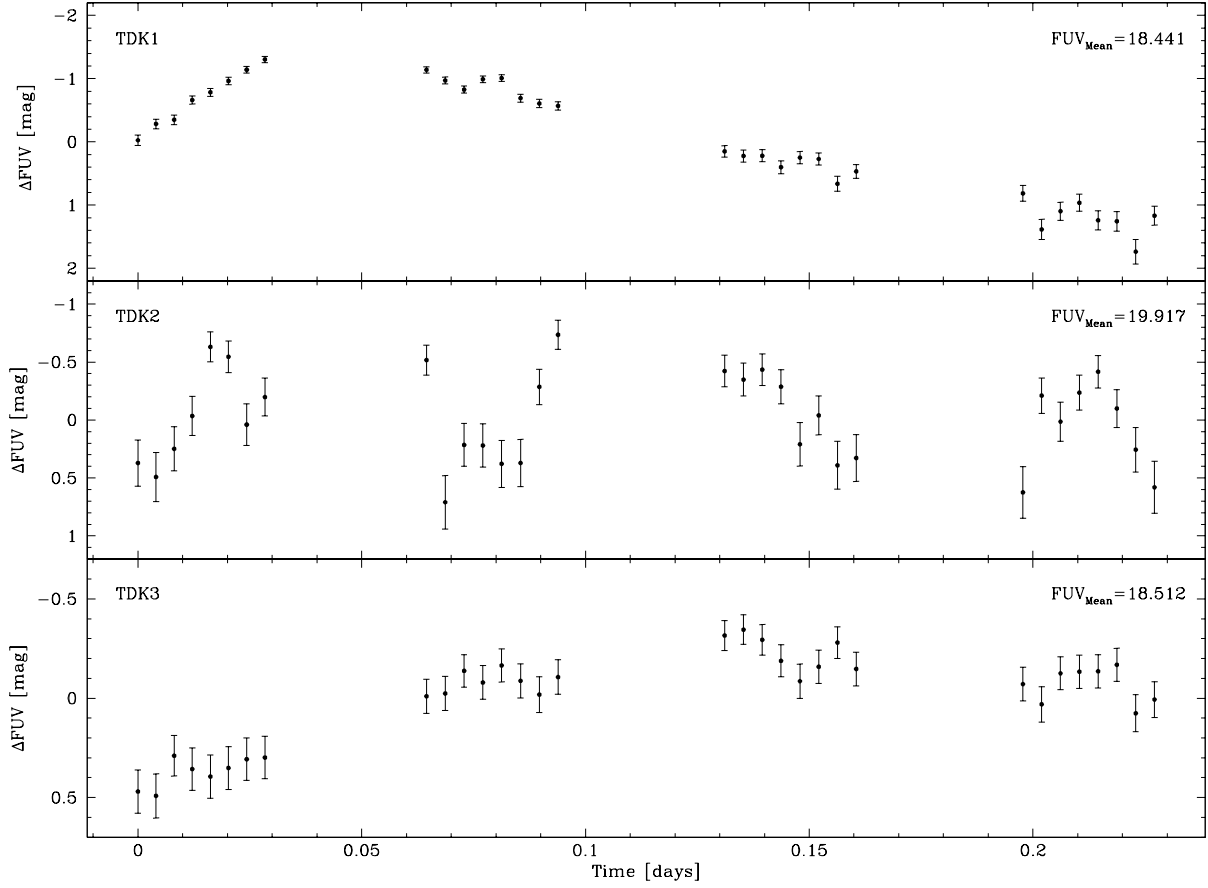}
\caption[Light curves of Other Variables]{Light curves for all the variable sources identified in M\,80, showing the variation from the mean magnitude ($\Delta FUV=FUV-FUV_{Mean}$). All light curves are detrended to remove trends due to the period of HST's orbit which are visible in the light curves of the brightest sources. Source TDK\,2 has a period of $\approx 55$ minutes and is likely an SX\,Phoenicis star. TDK\,3 shows long-term variability and might be an RR\,Lyrae or a Cepheid. See text for details.}
\label{fig_lightcurve_3_vars}
\end{figure*}

\begin{figure*}
\includegraphics[angle=270,width=0.8\textwidth]{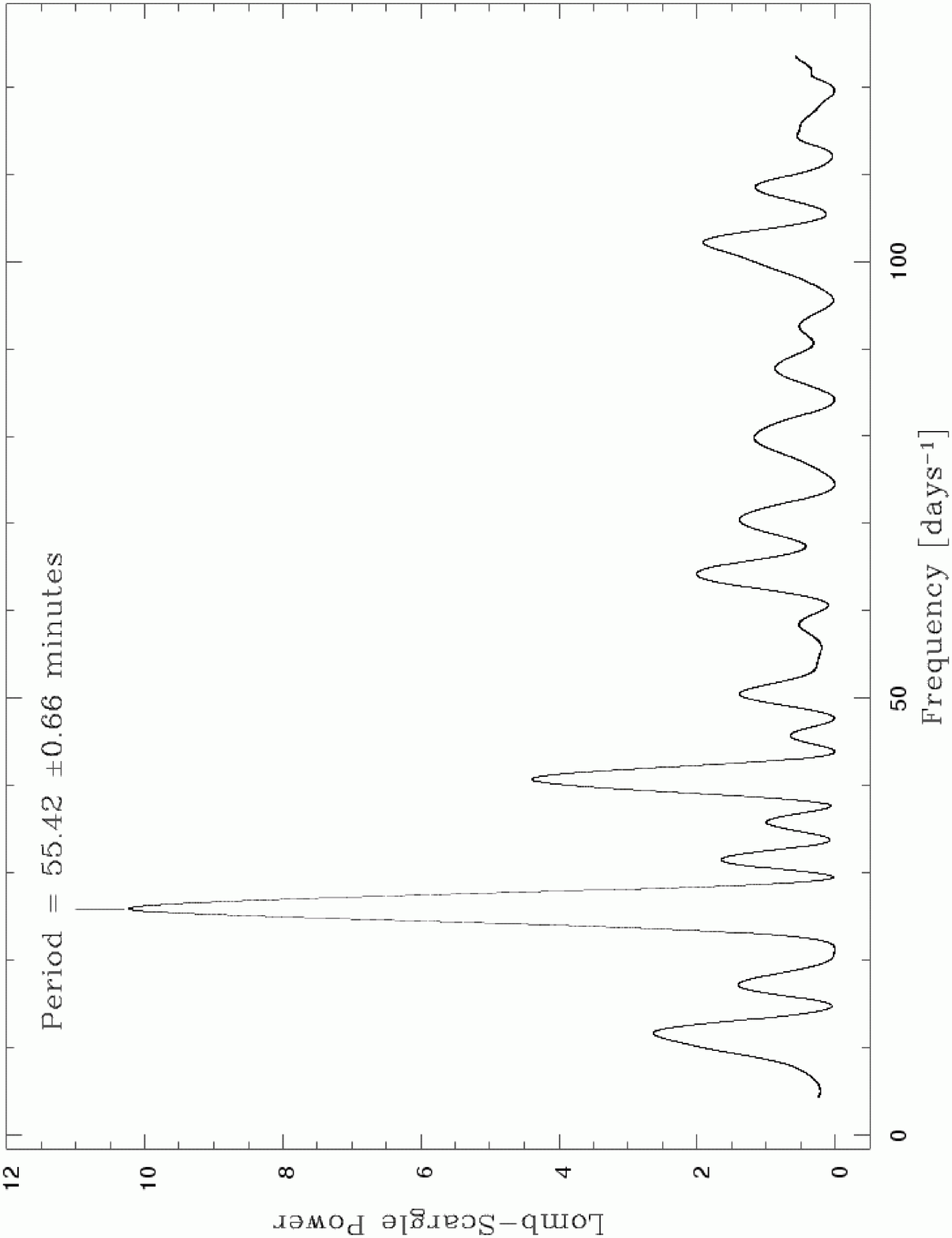}
\caption[Periodogram]{Lomb-Scargle periodogram of TDK\,2. The strongest peak corresponds to the period of 55.418 minutes.}
\label{fig_periodogram_2238}
\end{figure*}

\begin{figure*}
\includegraphics[angle=270,width=0.8\textwidth]{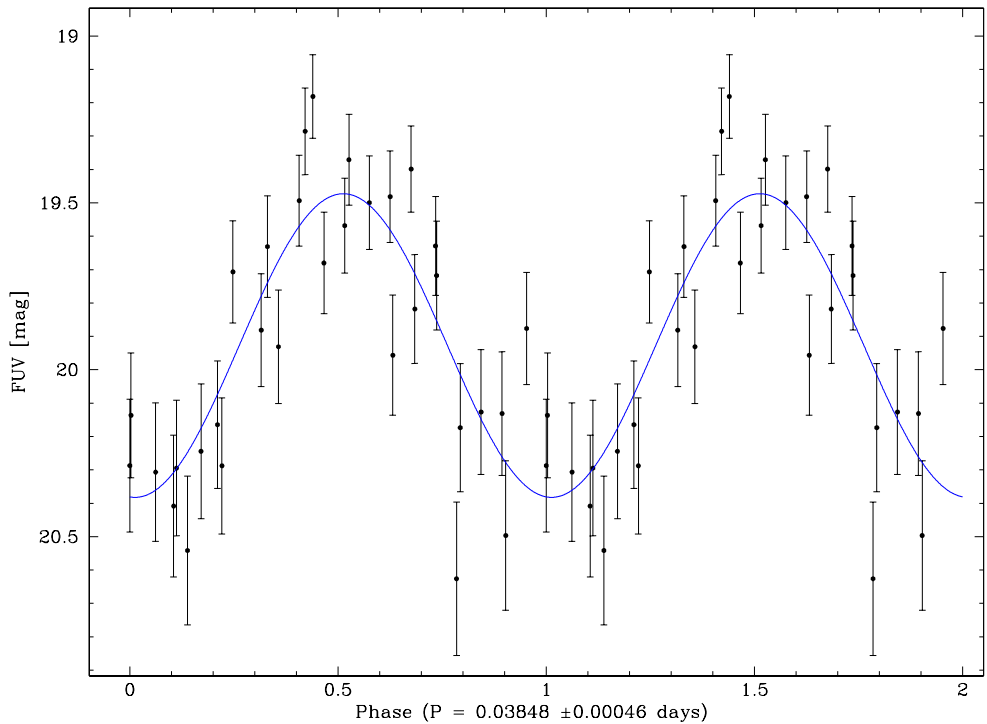}
\caption[Folded light curve 2238]{Light curve of variable source TDK\,2, folded with a period of 0.03848 days, or 55.418 minutes. Two complete cycles are shown. The blue line is a sinusoidal fit with the same period and an amplitude of 0.45 mag.}
\label{fig_folded_lightcurve_2238}
\end{figure*}

The light curve of TDK\,2 (source no.~2238 in Paper~1; see Figure~\ref{fig_lightcurve_3_vars}) shows short-term variability with a peak to peak variation of $\approx 1$ mag. Figure~\ref{fig_periodogram_2238} presents the Lomb-Scargle periodogram, which suggests a period of $55.42\pm0.66$ minutes. Figure \ref{fig_folded_lightcurve_2238} displays the light curve of TDK\,2, folded on this period, along with a sinusoidal fit which suggests a semi-amplitude of 0.45 mag. The position of TDK\,2 is in the blue straggler (BS) region in both the UV and optical CMDs (Figure~\ref{fig_cmd}). This, together with the period of $\approx1$ hour, suggests that TDK\,2 is an SX\,Phoenicis star.

SX\,Phoenicis stars are pulsating BS stars with periods of $\lesssim2$ hours and optical amplitudes of $\approx0.7$\,mag. They are Population II stars and are far more commonly found in GCs than in the field (Rodr\'{\i}guez \& L\'opez-Gonz\'alez 2000; Jeon et al. 2001). They are located in the BS region or the lowest part of the instability strip on the CMD, in the region associated with (Population I) $\delta$\,Scuti stars, so are thought to be their low metallicity counterparts. The physical characteristics of SX\,Phoenicis stars are not well explained by current theory, and the origin of BS stars is still a topic of discussion (see e.g., Knigge et al. 2009).

Further observations would be needed to determine additional information about TDK\,2. The short time span over which the observations were taken in our survey limit our ability to determine the period to a high level of accuracy, or detect the presence of multiple pulsation modes. If a pulsation spectrum could be obtained, parameters such as mass and metallicity may be found and could lead to new insights into the formation and evolution of SX\,Phoenicis stars and BS stars in general.

\subsubsection{TDK\,3}

The light curve of our source TDK\,3 (source no.~2324 in Paper~1), shown in Figure~\ref{fig_lightcurve_3_vars}, displays variability with $\gtrsim0.5$ mag semi-amplitude, with a period significantly longer than the timespan of our FUV observations. Our data coverage is not good enough to attempt a period determination for such long-term trends. Based on its position in the CMDs (Figure~\ref{fig_cmd}), however, TDK\,3 is likely to be another RR\,Lyrae star or a Cepheid variable. The shallow rise in brightness makes a type~ab RR\,Lyrae scenario unlikely, but TDK\,3 could be a type~c RR\,Lyrae star. RRc stars have shorter periods than RRab ($\lesssim0.5$ days) and FUV amplitudes of 1.3-3.3 mag (Downes et al. 2004), while Cepheid variables have periods on the order of days, with FUV amplitudes of around 1 magnitude (Smith et al. 2005; Moffett et al. 1998). Both of these are plausible explanations for the lightcurve of TDK\,3; our data does not allow us to constrain the parameters of TDK\,3 enough to determine which is the most likely option.

\subsection{Previously Known Variable Sources}

In addition to the sources described above which were identified in our variability search of the FUV data, we also discuss the FUV counterparts to two well-known variable sources, the classical nova T\,Sco and the dwarf nova DN\,1.

\subsubsection{T\,Scorpii}

Source number 2129 in our FUV catalogue was identified by us (see Paper~1) as the counterpart to the classical nova T\,Scorpii. This source was not strongly variable in our FUV observations. In fact, after subtracting trends due to PSF changes over HST's orbital period that were present in the light curves of the brightest sources, T\,Sco exhibits only very small scale variability in the FUV data, with amplitude $<0.1$ magnitudes. Bruch (1992) suggests nova flickering amplitudes of a few tenths of a magnitude and one would expect more variation in the bluer wavebands, but this was not detected. As described in Paper~1, T\,Sco was among the brightest objects in our FUV catalogue (FUV $= 15.44 \pm 0.01$ mag). It was also the bluest object in the catalogue; in fact, it was found to be unphysically blue ($\textrm{FUV}-\textrm{NUV}=-3.81$ mag; an infinite temperature blackbody would have $\textrm{FUV}-\textrm{NUV}=-1.8$ mag), so must have decreased in brightness in the month between the FUV and NUV observations, implying that T\,Sco was in a high state during the FUV observations. We suggest, therefore, that the flickering normally observed in the light curves of classical novae might be suppressed because we caught the source in a high state. The implied FUV outburst is interesting, as we cannot rule out the possibility that T\,Sco is, in fact, a recurrent nova, although this is probably a far-fetched explanation for the inferred FUV brightening; a DN eruption, for example, is a much more likely scenario.

\subsubsection{DN\,1}

Source number 1387 in our catalogue is the relatively faint FUV counterpart to X-ray source CX07 identified by Heinke et al. (2003). This source is also DN1, one of the DNe found by Shara \& Drissen (1995). Again, this source did not exhibit detectable variability in our FUV data. However, DN\,1 is a relatively faint FUV source (FUV$=22.578$), so instrumental errors are large ($\lesssim0.9$ mag) and limit our ability to draw any conclusions about the presence or absence of flickering on orbital variations.

\section{Summary}
\label{summary}

We used 32 individual FUV images from our UV survey of the core region of M\,80 (Dieball et al. 2010; referred to throughout this paper as Paper~1) to search for variable sources in our FUV catalogue. Three sources exhibit strong evidence for variability.

TDK\,1 (Paper~1's source no. 2817) is an RR\,Lyrae in the core of the cluster. The FUV light curve shows that it was observed from around 40 minutes before to 4.5 hours after maximum brightness, and further investigation using archival WFPC2 optical data showed that it is clearly variable in all wavebands. Its SED is reasonably well described by a star of temperature $T_{eff}\approx6700$\,K and radius $R\approx4.2\,R_{\odot}$, consistent with expected parameters for an RR\,Lyrae star. More specifically, we show that TDK\,1 is a type~ab RR\,Lyrae, based on the asymmetry in the FUV light curve.

This is only the third cluster in which an RR\,Lyrae star has been identified based on UV observations (others were found in NGC\,1851 by Downes et al. (2004) and M\,15 by Dieball et al. (2007)). UV surveys can be useful tools in identifying RR\,Lyraes and similar objects, particularly in the cores of (dense) GCs where optical surveys are seriously hampered by crowding.

TDK\,2 (source no.~2238 in Paper~1) is likely an SX\,Phoenicis star with a period of $55.42 \pm 0.66$ minutes and amplitude of $\approx1$\,mag. TDK\,3 (source no.~2324) might be another RR\,Lyrae or a Cepheid.

Finally, we discussed two well known variable sources, T\,Sco and DN\,1, the FUV counterparts of which were recovered in our survey. T\,Sco exhibited surprisingly little flickering in our FUV data, possiby because it was in a high state compared with the NUV observations a month later. DN\,1 is a very faint UV source, so photometric errors dominate over any possible intrinsic flickering or other variations.

After this paper was completed, we found that TDK\,1 and TDK\,3 are included in Kopacki's variability survey of M80 (Kopacki, private communication). Kopacki agrees with our classification of these two sources as RR\,Lyrae stars. A preliminary summary of his results, including periods but not including coordinates or finder charts for the sources, is given in Kopacki (2009).

\section*{Acknowledgments}

We thank Tom Maccarone, Brian Warner, Tom Marsh and Peter Schneider for helpful discussions. This work was supported by NASA through grant GO-10183 from the Space Telescope Science Institute, which is operated by AURA, Inc., under NASA contract NAS5-26555. A portion of this work was carried out at the Kavli Institute for Theoretical Physics in Santa Barbara CA, USA. This research was supported in part by the National Science Foundation under Grant No. PHY05-51164.

\newpage

\label{lastpage}
\end{document}